# Arbitrary tunable spaser based on double-Fano resonance of two sets of disk-ring nanostructure

Y. Huo,[1,a] T. Jia[2], C. Zheng[2], H. Zhao[2], S. Jiang[1], T. Ning[1], C. Yang[1], C. Tan[1], Y. Jiao[1], B. Man[1,b]

[1]*School of Physics and Electronics, Shandong Normal University, Jinan 250014, P.R. China*

[2]*State Key Laboratory of Precision Spectroscopy, Department of Physics, East China Normal University, Shanghai 200062, P.R. China*

This paper demonstrates an arbitrary tunable spaser based on double-Fano resonance of a plasmonic nanostructure consisting of two sets of disk-ring (TSDR) nanostructure. TSDR nanostructure supports two Fano resonances, which can be served as the lasing mode and the pumping mode of a spaser. These two mode can be tuned in a very wide wavelength range because of the good tunability of the two nanorings. The tuning range of the lasing mode and the pumping mode can be reached to 710 nm and 620 nm. These results represent a significant step in the pursuit of ultimate spasers and propose a approach to manipulate lasing mode and pumping mode over a broad spectral range.

Since spaser (surface plasmon amplification by stimulated emission of radiation) was proposed in 2003 by Bergman and Stockman,[1] it has been studied theoretically and verified experimentally.[2-4, 7-11] Spaser supports ultrasmall sizes below the diffraction limit and shows ultrafast dynamics, which is promising to be nanoscale coherent sources of optical fields.[1] It consists of noble metal nanostructures and gain materials.[5, 6] Metal nanostructures support surface plasmon (SP) modes as optical feedback, amplified by the resonant energy transfered from the gain materials. Gain materials have spectral and spatial overlap with the SP mode in order to compensate for intrinsic losses and amplify the SP mode of

[a] yanyanhuo2014@sdnu.edu.cn

[b] byman@sdnu.edu.cn

the metal nanostructures.[6] Early work of the spaser primarily focused on suppressing loss and amplifying surface plasmons through new nanostructure designs and materials. [4, 7-11]

Recent two years, tuning of lasing emission wavelength of plasmonic nanolasers attracted more extensive attention.[9, 12-14] R. Ma reported multicolored plasmon light sources by changing the width of the nanocavity and by tuning the electron−hole pair concentration in gain materials through direct electrical modulation;[12] Y. Lu studied the nanorod emitter tuned from blue to red across the visible spectrum by changing the composition of the gain materials;[9] A. Yang realized a real-time tunable lattice plasmon lasing based on arrays of gold nanoparticles and liquid gain materials;[13] X. Meng changed the gain material concentration to tune the resonance wavelength of the spaser.[14] Although different reports have demonstrated different wavelengths, a approach to manipulate lasing mode over a broad spectral range has been lacking.

The metallic disk-ring is a highly tunable nanostructure,[15] which can support multipolar Fano resonance.[16-19] Y. H. Fu found that higher-order Fano resonances in the dual-disk ring nanostructure.[16] Y. Zhang proposed a disk-ring nanostructure to produce Fano resonance with narrow line-width and high CR, and show that sharp multipolar Fano resonances can be induced.[17, 18] S. Liu studied multiple Fano resonances in plasmonic heptamer clusters composed of split nanorings split nanorings clusters.[19] Fano resonance can excite the dark mode surface plasmon, which can be amplified to realized spaser.[20-22] Spaser based on Fano resonance supports small mode volume, lower threshold and higher Purcell factor.[22] The Fano resonance wavelength of the disk-ring nanostructures mentioned above can be arbitrarily tuned by adjusting the structure parameters, which lays the foundation for realizing an arbitrary tunable spaser based on two dark mode surface plasmons. Spaser based on two resonances can enhance the pumping efficiency and reduce the energy threshold.[23] However, it must ensure that the two resonances have common spatial overlap field.[6] Only then the gain materials can be pumped to the excited state by one resonance, and the excited state population can be stimulated downward to the lower state by another resonance.

In this paper, we proposed a nanostructure consisting of two sets of disk-ring (TSDR) nanostructure, which can supports two Fano resonances. The two Fano resonance wavelengths not only can be adjusted arbitrarily, but also their electric field distribution have common spatial overlap. We study the spaser based on the two Fano resonances supported by TSDR nanostructure and the arbitrary tunability of this spaser's lasing mode and pumping mode.

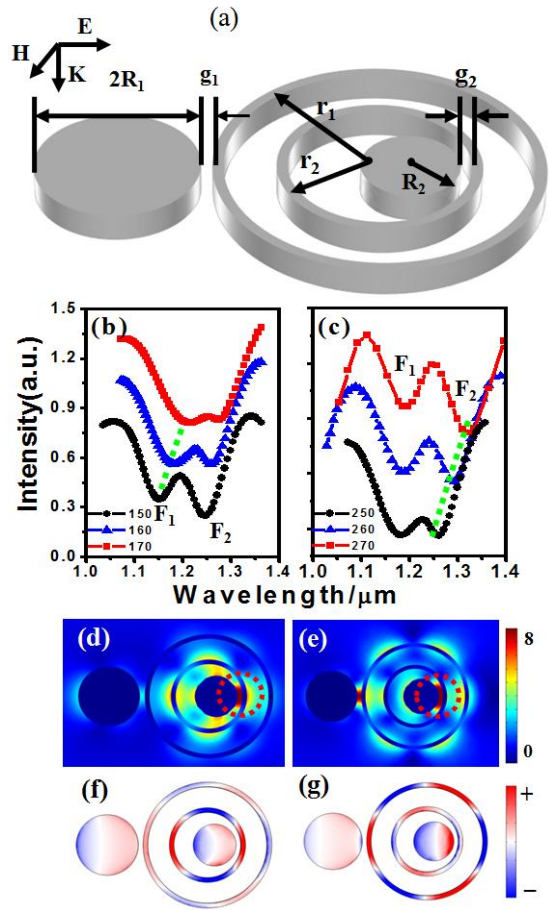

Fig. 1. (a) Sketch of the TSDR nanostructure and the incident light. The scattering spectra of TSDR nanostructure with different radius of inner ring (b) and outer ring (c) in the fused silica with no gain media, the radius of outer disk $R_1$=130 nm and inner disk $R_2$=90 nm. The electric field amplitude in the middle section (d) (e) and the induced surface charges on the top surfaces (f) (g) at $F_1$ Fano resonance (d) (f) and $F_2$ Fano resonance (e) (g).

The TSDR nanostructure of the spaser is schematically shown in Fig. 1 (a), which consists of an outer nanodisk with diameter $R_1$, an inner nanodisk with radius $R_2$, and two concentric nanorings. The radius of the outer nanoring and inner nanoring is $r_1$ and $r_2$, respectively. The gap between the outer nanodisk and the outer nanoring is $g_1 = 20$ nm. The inner nanodisk and the inner nanoring is decentered with a gap $g_2 = 20$ nm. In this paper, the nanoring wall width is 20 nm, and the height of all nanostructure keeps at 60 nm. The nanostructure is packed by fused silica with the refractive index $n_0 =$ 1.46. When the fused silica layer doped with gain materials, we assume that the refractive index of the gain layer is $n_g = 1.46 - ik$. $k$ is gain coefficient, which is related to the amplification coefficient of light intensity via $g = 4\pi k/\lambda$.[24] The materials of the metallic nanostructure is Ag, with the permeability $\mu = 1$ and the complex permittivity sourced from Refs. 25. We use the finite element method (COMSOL) to investigate the interaction between the surface plasmon of the TSDR nanostructure and the gain materials. A plane wave irradiates down to the TSDR nanostructure, and the electric field is parallel to the linked line of the centers of two nanodisks, as shown in Fig. 1 (a). For this polarization, this structure supports very strong Fano resonance.

Fig. 1 (b) shows the scattering spectra of the TSDR nanostructure with no gain media in the fused silica layer. There are two evident Fano resonances $F_1$ and $F_2$ in Fig. 1 (b) and (c). The electric field amplitude in the middle section and the induced surface charges on the top surfaces at $F_1$ and $F_2$ Fano resonance are shown in Fig. 1 (d-g), respectively. These show that $F_1$ Fano resonance origins from the interaction between the inner nanoring quadrupolar dark mode and the inner nanodisk dipole resonance. $F_2$ Fano resonance origins from the interaction between the outer nanoring octupolar dark mode and the outer nanodisk dipole resonance. So $F_1$ Fano resonance wavelength is mainly controlled by the inner set of disk-ring nanostructure, and $F_2$ Fano resonance wavelength is mainly controlled by the outer set of disk-ring nanostructure. The relative wavelength position and absolute wavelength position of these two Fano resonances can be tuned arbitrarily by adjusting the two nanoring, respectively. Fig. 1 (b) and (c) show the tunability of the TSDR nanostructure. In Fig. 1 (b), we keep the radius of the outer nanoring $r_1$ = 250 nm unchanged, adjust the inner nanoring radius from 150 nm to 170 nm, $F_1$ Fano resonance wavelength can be tuned from 1149 nm to 1217 nm. When we keep the size of the inner ring $r_2$ = 160 nm unchanged, adjust the radius of the outer nanoring from 250 nm to 270 nm, $F_2$ Fano resonance wavelength can be tuned from 1181 nm to 1321 nm, as shown in Fig. 1. (c). This TSDR nanostructure not only has a very good tunability, but also its two Fano resonances have common spatial overlap, as the red dotted circle shown in Fig. 1. (d) and (e). So this TSDR nanostructure is right for realizing an arbitrary tunable spaser based on two dark mode surface plasmon resonances.

According to the absorption peak and emission peak of $Er^{3+}$ ion, we carefully adjust the two nanorings and nanodisks radius. It's easy to make two Fano resonances close to 980 nm absorption peak and 1550 nm emission peak of $Er^{3+}$ ion, as shown with blue arrow in Fig. 2 (a). As the electric field distribution in Fig 2 (a) insets shown, we can see that 980 nm Fano resonance origins from the inner nanoring octupolar resonance, 1550 nm Fano resonance origins from the outer ring octupolar resonance. When we select $Er^{3+}$ ion as the gain material, the 980 nm and 1550 nm Fano resonance can be served as the pumping mode and the lasing mode, respectively. Fig. 2 (b) shows the dark mode excited by 1550 nm Fano resonance is amplified by stimulated emission of radiation as the lasing mode. When gain coefficient $k$ reach to the threshold 0.08515, the scatter peak value of outer nanoring octupolar resonance can reach to $2.7 \times 10^7$, which is $1.03 \times 10^8$ times higher than the scatter peak of the TSDR nanostructure with no gain material. The linewidth is significantly compressed. The energy provided by gain materials for the lasing mode amplification and the system loss keeps a dynamic balance. The dark mode surface plasmon at the 1550 nm Fano resonance gets the best amplification. In Fig. 2 (a), there are two Fano resonances appear at 1167 nm and 1357 nm, which origin from the outer nanoring hexadecapolar resonance and the inner nanoring quadrupolar resonance, respectively.

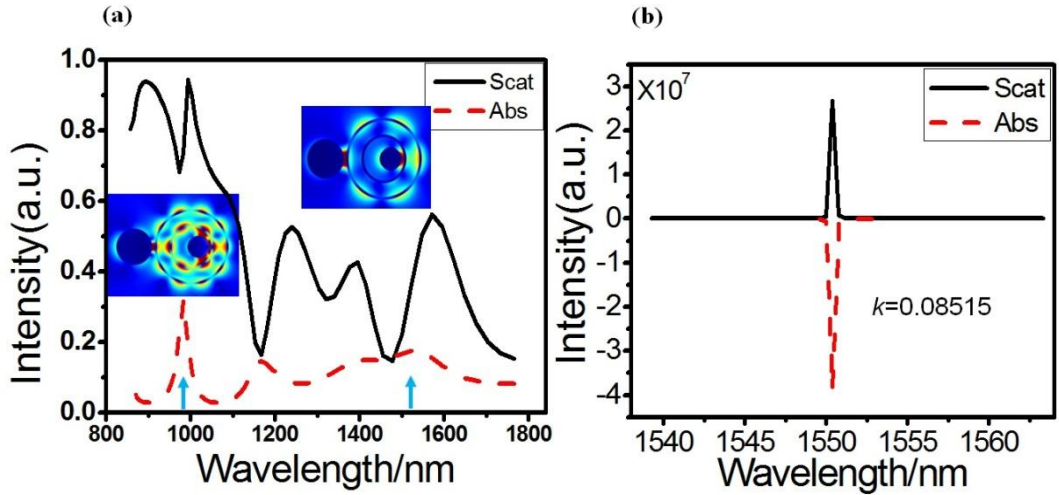

Fig. 2 (a) The scattering spectra and absorption spectra of TSDR with $R_1 = 150$ nm, $R_2 = 90$ nm, $r_1 = 310$ nm, $r_2 = 185$ nm. Inset: the electric field amplitude in the middle section at 980 nm Fano resonance (below left) and 1550 nm Fano resonance (up right). (b) The scattering spectra and absorption spectra of TSDR nanostructure with gain coefficient $k$ at the threshold.

We keep the inner nanoring octupolar Fano resonance wavelength, i.e. the 980 nm pumping mode unchanged, adjust the radius of the outer nanoring from 260 nm to 350 nm, the lasing mode only decided by the outer nanoring octupolar resonance can be tuned from 1.32 μm to 1.71 μm, as shown in Fig. 3 (a). In addition, the quadrupolar Fano resonance of the outer nanoring can also be considered as the lasing mode. In this way, the lasing mode wavelength can be tuned from 1.0 μm to 1.71 μm, the tuning range can be expanded to 710 nm. If we don't consider the limit of Ag dielectric constant, we can further increase the radius of the outer nanoring, the lasing wavelength can be tuned to the middle infra-red wavelength ranges, even to the far infra-red wavelength ranges. Fig. 3 (b) shows the lasing modes decided by the octupolar or hexadecapolar Fano resonance of the outer nanoring are amplified by stimulated emission of radiation. We can see every dark mode can be amplified, just the threshold of every lasing mode is different, which labeled in Fig. 3 (b). The threshold of lasing mode based on the octupolar Fano resonance is generally higher than the lasing mode based on the hexadecapolar Fano resonance.

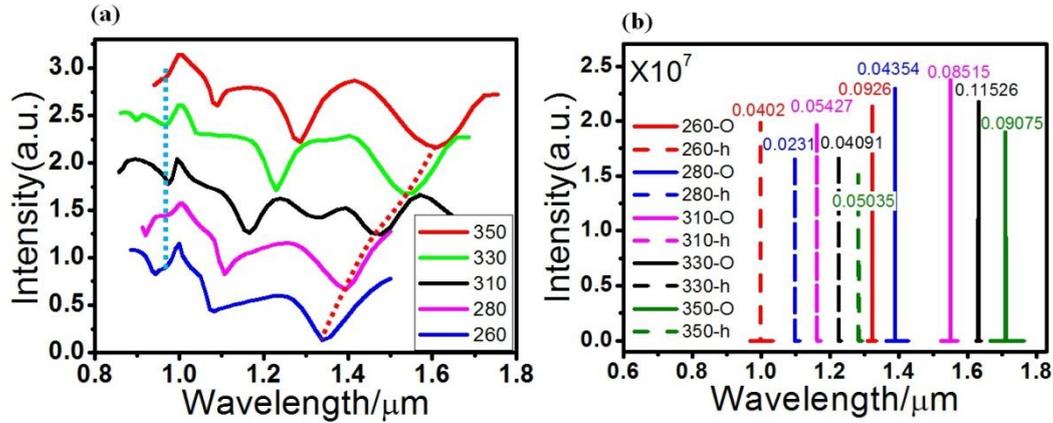

Fig. 3 (a) The scattering spectrum of TSDR nanostruture with no gain material in fused silica change as the outer nanoring radius. (b) The scattering spectra of octupolar (solid lines) and hexadecapolar (dashed lines) dark mode amplified by stimulated emission of radiation at the threshold *k*. The threshold of every mode is labeled in figure (b) .

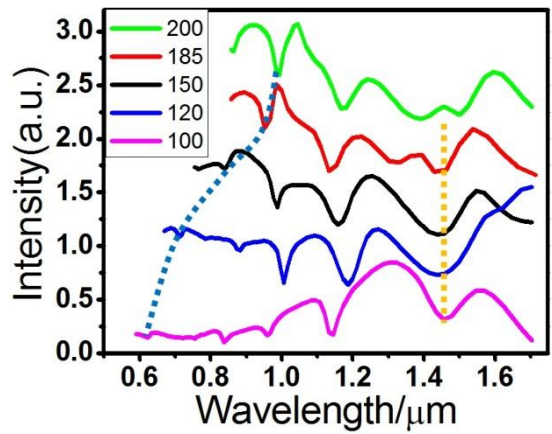

Fig. 4 The scattering spectrum of TSDR nanostructures with no gain material in fused silica change as the inner nanoring radius.

Fig. 4 shows that keeping the outer nanoring octupolar Fano resonance wavelength, i.e. the 1550 nm lasing mode unchanged, adjust the radius of the inner nanoring from 100 nm to 200 nm, the pumping mode decided by the inner nanoring octupolar resonance can be tuned from 0.62 μm to 1.24 μm, the tuning range also can be expanded to 620 nm.

In summary, we theoretically realize an arbitrarily tunable spaser based on double-Fano resonance of a TSDR nanostructure. The TSDR nanostructure can support two Fano resonances, which can be controlled by the two nanorings respectively. The two Fano resonances are served as the lasing mode and pumping mode of the spaser. The tuning range of the lasing mode and pumping mode can be reached to 710 nm and 620 nm. Spaser based on two Fano resonances not only supports small mode volume, lower threshold and higher Purcell factor, but also can enhance the pumping efficiency and reduce the energy threshold. These results represent a significant step in the pursuit of ultimate spasers and propose a approach to manipulate lasing mode and pumping mode over a broad spectral range.

**Acknowledgments**

The authors are grateful for financial supports from the National Natural Science Foundation of China (11504209, 11404195, 61205174, 11474187, 11405098), Youth Science and Technology Project Fostering Fund of Shandong Normal University, and Excellent Young Scholars Research Fund of Shandong Normal University.